\documentstyle[aps,12pt,manuscript,epsf]{revtex}
\begin{document}
\preprint{INJE--TP--96--3 }

\title{ Comment on ``Background thermal contributions in testing the Unruh effect''}

\author{ D.K.Park$^1$, H.W. Lee$^2$, Y. S. Myung$^2$ and Jin Young Kim$^3$}
\address{$^1$Department of Physics, Kyungnam University, Masan 631-701, Korea\\
         $^2$Department of Physics, Inje University, Kimhae 621-749, Korea\\
         $^3$Division of Basic Science, Dongseo University, Pusan 616-010, Korea} 

\maketitle
\vskip 1.5in

\begin{abstract}
Costa and Matsas claimed in their recent paper that a thermal bath does not
affect substantially the transition probability for fast moving inertial
Unruh detector. It is shown that their claim holds only for small 
$\beta \Delta E$.  We show that, for large enough $\beta \Delta E$, the 
transition probability is not monotonically decreasing function of the 
detector's speed contrary to their claim.  
\end{abstract}

\newpage
In a recent paper\cite{Costa}, Costa and Matsas investigated the background 
thermal contributions in
Unruh effect\cite{Unruh}.   They claimed that the faster a particle detector 
moves in a background 
thermal bath, the less the detector interacts with the bath  
and concluded that the thermal bath does not affect substantially the 
transition probability of the detector.  
Their claim is based on the following observation.  Time dilation induces a 
fast moving
detector to interacts preferentially with low energy modes.  
Although a thermal bath is rich of low frequency modes, the phase space 
volume element ($\propto \omega^2 d \omega$)
suppresses the infrared contributions.

In this comment we will show that their claim is true when $\beta \Delta E$ 
is comparatively
small, where $\beta = 1/ k T$ and $\Delta E = E - E_0$ is the energy 
difference of two level
detector.   When $\beta \Delta E$ is large enough, background thermal 
contribution would be 
dominant for the fast moving detector.  In order to show this let us start 
with Eq.(6) of Ref.[1]
\begin{equation}
{ {\cal P}^{exc} \over T^{tot} }  = { c_0^2 \sqrt{1-v^2} 
													\over 4 \pi v \beta}
       \ln \left [ {     1 - e^{ - \beta \Delta E \sqrt{1 + v} / 
															\sqrt {1 - v} }  
                  \over 1 - e^{ - \beta \Delta E \sqrt{1 - v} / 
											\sqrt {1 + v} }   } \right ],
\end{equation}
where ${\cal P}^{exc} / T^{tot}$ is the total transition rate per total 
proper time $T^{tot}$ of the detector. 
As commented in Ref.\cite{Costa} the $v \to 0$ limit of Eq.(1) recovers 
the Planckian 
spectrum
\begin{equation}
\lim_{v \to 0} { {\cal P}^{exc} \over T^{tot} }  = { c_0^2 \over 2 \pi }
       {\Delta E \over e^{\beta \Delta E} - 1}.
\end{equation}
The $v$-dependence of ${\cal P}^{exc} / T^{tot}$ is shown in Fig.1 for 
various values of 
$\beta \Delta E$\cite{Com}.  From Fig.1 one can see that for large enough 
$\beta \Delta E$, there always 
exists a speed $v_m$ such that the transition probability  
is maximized
if the detector moves with that speed. 
Numerical calculation provides that the condition for the existence of 
$v_m$ is $\beta \Delta E \ge 2.593$.  
The appearance of peaks in Fig. 1 means that the phase space 
volume element does not completely suppress the infrared contributions 
when $\beta \Delta E$ is large enough.
So the claim of Ref.\cite{Costa} is correct only when 
$\beta \Delta E < 2.593$, but is incorrect for $\beta \Delta E \ge 2.593$.  
Table I shows 
that $v_m$ approaches to the speed of light as $\beta \Delta E $ increases.  
Therefore, for large enough $\beta \Delta E $, the transition probability 
is not monotonically decreasing function of $v$ contrary to their claim. 
In addition, from Eq.(2) one can understand that the spectral density (
$(\Delta E)^2 {\cal P}^{exc} / T^{tot}$)
is a Doppler effect of Planckian spectrum.  This is plotted in Fig.2 for 
various values of
$v$.  As the speed of the detector increases, the peak becomes higher 
and moves 
to right as shown in Fig.2.  Note the similarity between Fig. 2 and 
the conventional Planck distribution.  
In conventional Planck distribution, which corresponds to $v=0$ for 
our case, the peak shifts to the right as the 
temperature increases.  Fig. 2 shows the same phenomena as velocity 
increases.  
This is because the relativistic 
temperature depends on $v$ as 
\begin{equation}
\beta \propto \sqrt{1-v \over 1 +v},
\end{equation}
which is discussed in Chapter 4 of Ref.\cite{Birrell}. 

\acknowledgments

This work was supported in part by Nondirected Research Fund, Korea 
Research Foundation,1994 and by Korea Science and Engineering 
Foundation (94--1400--04--01--3 and 961--0201--005--1).

\begin{figure}
\epsfysize=8cm \epsfbox{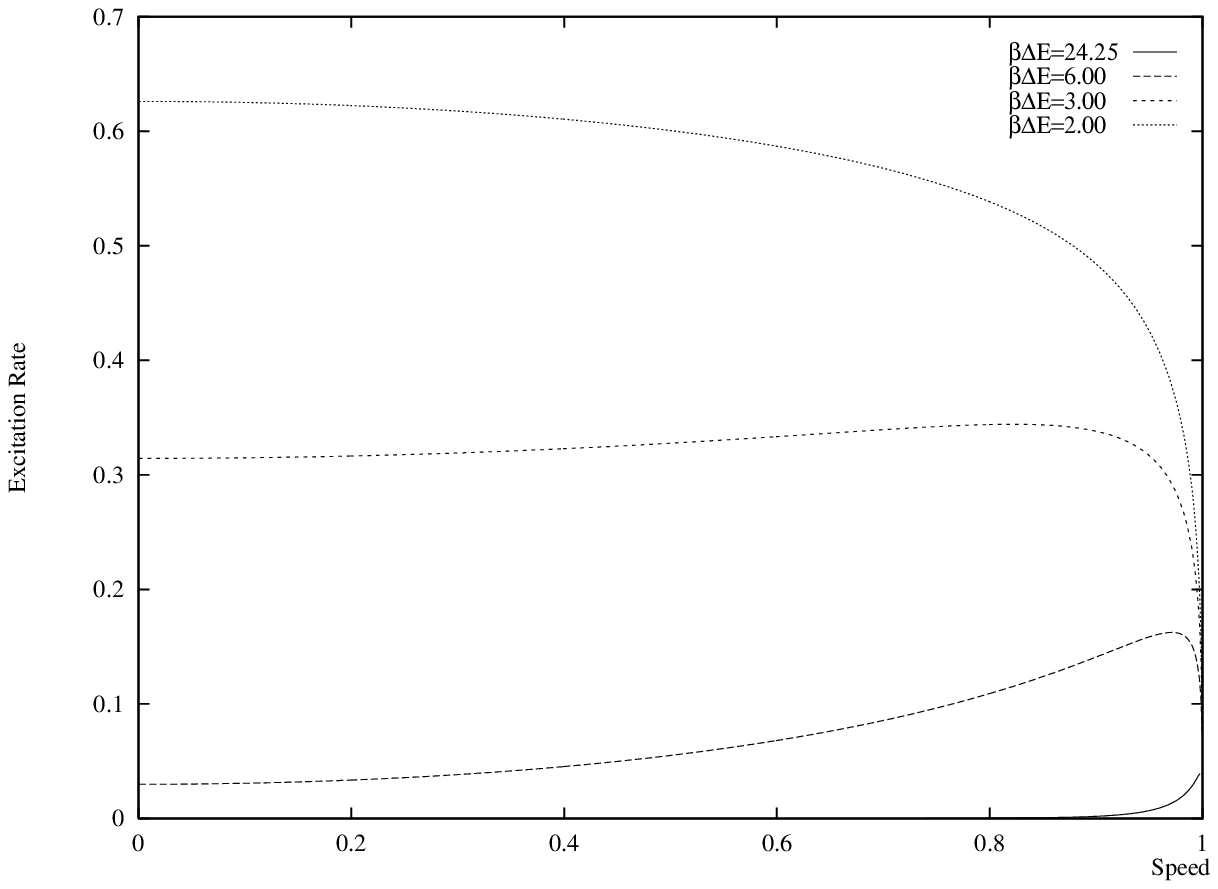}
\caption{$v$ dependence of excitation rate for various $\beta \Delta E$}
\end{figure}
\begin{figure}
\epsfysize=8cm \epsfbox{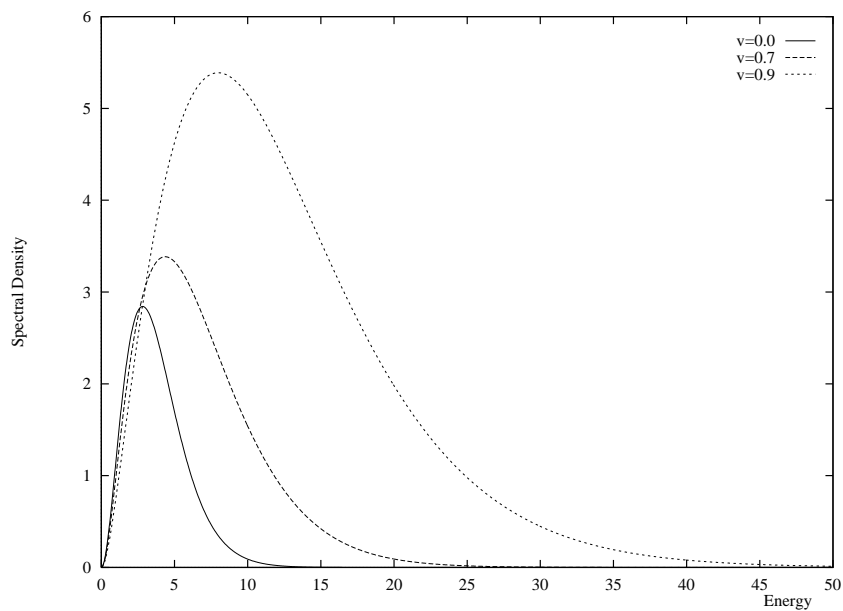}
\caption{Energy dependence of spectral density for various speeds}
\end{figure}

\begin{table}
\caption{Numerical values of $v_m$ for selected $\beta \Delta E$}
\begin{tabular}{cccccc}
$\beta \Delta E$ & 4& 6& 8& 10& 20 \\
\tableline
$v_m$ & 0.926059& 0.971074& 0.984319& 0.990125& 0.997582
\end{tabular} 
\end{table}


\begin{references}
\bibitem{Com} The $v$--dependence of ${\cal P}^{exc}/T^{tot}$ for 
$\beta \Delta E=24.25$, shown in Fig. 1 of Ref. \cite{Costa}, seems wrong 
compared with our result for the same $\beta \Delta E$(see Fig. 1).
\end{references}
\end{document}